# Molecular Insights into the Electrowetting Behavior of Aqueous Ionic Liquid

*Sanchari Bhattacharjee, Sandip Khan* *

Department of Chemical & Biochemical Engineering

Indian Institute of Technology Patna

Patna, India- 801103.

*Corresponding author: E-mail: skhan@iitp.ac.in

**Abstract**

Molecular dynamics simulations were employed to study the wetting behavior of nanoscale aqueous hydrophilic and hydrophobic Imidazolium based ionic liquid (IL) droplets on a solid graphite substrate subjected to the perpendicular electric field. Imminent transformation in the droplet configuration was observed at E = 0.08 V/Å both for hydrophobic ILs [EMIM][NTF$_2$] and SPC/E water droplets. However, for hydrophilic, [EMIM][BF$_4$], the droplet wets asymmetrically within electric field E = 0.09 V/Å for lower weight percentages of ILs and E = 0.1 V/Å at a higher weight percentage of ILs (i.e., 50wt%), because of the strong interaction of ILs with the sheet. We explore the impact of electric-field through various parameters such as mass density distribution, contact angle, orientation, and hydrogen bonds.

**Keywords**: electro-wetting; contact angle; aqueous hydrophilic and hydrophobic Imidazolium ILs, molecular dynamic simulation

# 1. Introduction

Electro-wetting is the modulation in the wettability of a surface prompted by an external voltage within the surface and the liquid [1,2] or across the solid/liquid interface. Ionic liquids (ILs) which have unique physicochemical properties,[3] gain industrial significance, due to its development as a promising alternative medium for electrowetting. This phenomenon is exploited in various applications namely, variable fluid lens systems, electrowetting displays, programmable optical filters, paint drying, micro motors, electronic microreactors, liquid optics, or lab-on-a-chip devices.[4,5,6]

It is acclaimed that an electric field can produce a noticeable dipole effect in water [4,7], since ILs comprise of ions and counter ions, an electric field may have an enormous impact on this ion–counter ion cage.[8], the concept of *ion cage* formation arises when the central ion is enclosed by its adjacent counter ions of the first coordination shell, which results in the local charge ordering in ILs,[9]. For instance, Wang et al. [10] observed spatial heterogeneity (i.e., disordering and reordering) of ionic liquids [$C_{12}$MIM][$NO_3$] with the implementation of an external electric field which further, changes from heterogeneous to homogeneous, and lastly to nematic-like with the rise in an external electric field.

Also, electrowetting of aqueous ILs [11] has been performed to investigate their possible applications in microfluidic devices [12,13]. Thus aqueous solutions or water content plays an imperative role in electrowetting based applications in microfluidic devices.[6] Promising electro wetting-based RTILs would improve their function as environmentally friendly solvents in various industries.[14,15] Although the consequence of the electric field on the bulk structures of ILs has been investigated broadly [9,10,16]. Nonetheless, its impact on the surface structure of ILs is remains unclear. Li and Restolho [2,17] stated that the electro wetting nature of ionic liquids was analogous to that of traditional saline solutions wherein decrement in the contact angle was found with an increase of the implemented voltage and the rise of the volume of ions was one of the cause which was held responsible for this decrement.

Suitable electrowetting-based RTILs would strengthen their role as environmentally friendly solvents in various industries.[14,15]. Though the effect of the electric field on the bulk structures of ILs has been studied broadly [9,10,16]. Nonetheless, its impact on the surface structure of ILs is still not clear. Li and Restolho [2,17] stated that the electrowetting nature of ionic liquids was analogous to that of traditional saline solutions wherein decrement in the contact angle was found with an increase of the implemented voltage due to increase in the volume of the ions.

Millefiorini et al.[15] was the first to demonstrate the asymmetric nature of the contact angles on the fluoropolymer surface for Imidazolium and Pyrrolidinium based ILs with the application of an external DC voltage. The asymmetric behavior rises when the contact angle reaches a limiting value (the contact angle value would not decrease following further increase of neither positive nor negative voltages), similar behavior of asymmetric contact angle profile for the electrode potential was also found by other authors [18-20]. The limiting point is dependent on specific ions size and structure. They additionally added water has a comparatively minor impact on the surface tension and contact angle measurement.

Interfacial and electrowetting behavior of ionic liquids with common cation [EMIM] associated with a different anion such as [TFSI], [TFO], [OMS], [OAC], [FSI], [DCA], and [FAP] on bare metallic electrodes were investigated by Zhen Liu et al. [21]. They revealed through AFM studies that the formation of precursor film at negative electrode tends to spread the IL at negative potentials, which is unlikely at positive potentials that become the reason of asymmetric behavior of ILs in electrowetting, additionally insertion of 10% of water to ILs has a very minor effect on the contact angle. The impact of water on the electrowetting application of the [BMIM][BF$_4$] in-phase of hexadecane was reported by Paneru et al.[13] , Their results showed neither ac nor dc potentials had any undesirable effects with the insertion of water over the whole concentration range of ILs (0 and 99%); moreover, changes in the interfacial tension and viscosity occurs, rather than a change in wettability. Similarly, when an immiscible IL is diluted with water, the contact angle has also been displaying no dependence on water concentration, irrespective of only altering thermodynamic properties [19,22].

Another exciting influence of the external electric field on IL is the phase change behavior, in which hydrophobic IL water systems changed from two-phase to single-phase systems.[23] Sha et al.[23] revealed hydrophobic to the hydrophilic transition of (octylmethylimidazolium hexafluorophosphate) [OMIM][PF$_6$] induced by an electric field both from atomic force microscopy (AFM) experiments and non-equilibrium molecular dynamics (MD) simulations. Despite extensive study of bulk RTIL performance at charged surfaces through experiments and simulations study, [28-32] the electro wetting states of RTIL nano-droplets persist in scares.

Taherian et al. [24] used MD simulation to calculate 2D contact angles of [BMIM][BF$_4$] on positive and negative charged graphene surfaces. They found an asymmetry in the electrowetting aspect of the IL. Additionally, the surface with negative charges indicated (lower contact angle) droplet in comparison to the positively charged surface, which is because of the orientation predilections of the cations in the three-phase contact line. Further, Song et al.[25]

reported asymmetrical wetting of 1-butyl-3-methylimidazolium tetrafluoroborate [BMIM][BF$_4$] on a silicon solid substrate on hydrophobic potential ($\varepsilon = 0.1$ kcal/mol). However, the asymmetrical wetting phenomenon mostly vanishes at hydrophilic potential ($\varepsilon = 2.0$ kcal/mol) due to the strong attraction of substrate towards ionic liquid particles. Thus, interaction within cationic/anionic particles with and substrates plays a vital role in wettability.

The above mention literature suggested that interfacial IL structure and addition of water plays has a substantial impact on the contact angle measurement. It is, therefore, crucial to unravel the correlation among the Ionic liquid interfacial structure and the electrowettability of ILs. The electrowetting of a droplet has been investigated experimentally for a macro-sized droplet [26]. However, the dynamics of a nanosized droplet, under the influence of an electric field, appears to have been elusive. Therefore, the dynamics of a nano-drop under the impact of an electric field is investigated employing MD simulations

Interfacial and electro wetting behavior of ionic liquids with common cation (EMIM) associated with a different anion such as [TFSI], [TFO], [OMS], [OAC], [FSI], [DCA], and [FAP] on bare metallic electrodes were investigated by Zhen Liu et al. [21]. They revealed through AFM studies that the formation of precursor film at negative electrode tends to spread the IL at negative potentials, which is unlikely at positive potentials that becomes the reason of asymmetric behavior of ILs in electrowetting, additionally insertion of 10% of water to ILs has a very negligible effect on the contact angle. The impact of water on the electrowetting application of the [BMIM][BF$_4$] in-phase of hexadecane was reported by Paneru et al.[13]. Their results showed neither ac nor dc potentials had any undesirable effects with the insertion of water over the whole concentration range of ILs (0 and 99%); moreover, changes in the interfacial tension and viscosity occurs, rather than a change in wettability. Similarly, when an immiscible IL is diluted with water, the contact angle has also been shown to have no dependence on water concentration, despite only altering thermodynamic properties [19,22].

Another exciting influence of the external electric field on IL is the phase change behavior, in which hydrophobic IL water systems changed from two-phase to single-phase systems.[23] Sha et al.[23] revealed hydrophobic to the hydrophilic transition of (octylmethylimidazolium hexafluorophosphate) [OMIM][PF$_6$] induced by an electric field both from atomic force microscopy (AFM) experiments and non-equilibrium molecular dynamics (MD) simulations. In spite of extensive study of bulk RTIL performance at charged surfaces through experiments and simulations study, [28-32] the electro wetting states of RTIL nano-droplets persist in scares.

The above mention literature suggested that interfacial IL structure and addition of water plays has a substantial impact on the contact angle measurement. It is, therefore, crucial to unravel the correlation between the interfacial IL structure and the electro-wettability of ILs. The electrowetting of a droplet has been investigated experimentally for a macro-sized droplet [26]. However, the dynamics of a nanosized droplet, under the influence of an electric field, appears to have been elusive. Therefore, the dynamics of a nano droplet on a surfaces under the impact of an electric field is investigated employing MD simulations. In this article, we present a molecular-level understanding of aqueous hydrophilic 1-Ethyl-3-methylimidazolium tetrafluoroborate [EMIM][BF$_4$] and hydrophobic 1-Ethyl-3-methylimidazolium bis(trifluoromethylsulfonyl)imide [EMIM][NTF$_2$] droplet on electrowetting behavior at graphite surface.

## 2. Computational Model and Methods

An equilibrated drop of aqueous ILs [EMIM][BF$_4$] and [EMIM][NTF$_2$] from our previous[27] publication is taken in this study. The Lennard-Jones interactions between graphite atoms and aqueous ILs are taken from Werder et al.[28] SPC/E model [29]of water is used to solvate the system

The force field used for ILs taken from Lopes et al. [30]Three layers of graphite surface were used with an interlayer spacing of 3.35 Å to cover the cutoff range. The graphene sheets were fixed during the simulation. To avoid interaction among periodic pictures of the droplet, the height of the simulation box was kept large enough (z dimension~200 Å). The cross interaction parameters between different atoms were estimated using Lorentz−Berthelot arithmetic mixing rules [31]. The cut-off distances for the nonbonded interactions were taken as 12Å. For Coulomb interactions cutoff of 12 Å, and 10 Å for LJ interaction was taken. The Particle-Particle Particle-Mesh (PPPM)[32] was used for the long-range electrostatic interactions with an accuracy of 0.0001.

**2.2 Simulation Systems**

An equilibrated drop was initially placed on the graphite surface, and periodic boundary conditions were used in three directions of the simulation box. The applied external electric fields had magnitudes of E = 0.0, +0.03, +0.05, +0.07, +0.08, +0.09 and +0.1 V/Å and were perpendicular to the solid substrate. All the systems were executed in NVT ensemble at 300 K for 20 ns using molecular dynamics software, LAMMPS [33]. The Nose−Hoover thermostat applied to maintain the temperature (298 K), and the velocity−verlet algorithm used to integrate Newton's equation of motion. Details of the contact angle measurement can be found in our previous publications [34].

## 3. Results and Discussion

### 3.1 Effect of Electric Field on Water Nano-droplet

Before discussing the electrowetting behavior of aqueous ionic liquid, first, we have investigated the behavior of pure water (SPC/E) droplets on the graphite sheet under the effect of the external electrical field. Snapshots of water droplets with 3000 molecules at 300 K and the corresponding variation of contact angle of the droplet on the graphite sheet at the various electric field are displayed in **Figure 1**. Snapshots and contact angle plots indicate that the wetting characteristics of water nanodroplets considerably rely on the intensity of electric fields. For example, the contact angle of the droplet is unchanged up to 0.05 V/Å. However, at 0.07 V/Å, a noticeable decrement of contact angle ($\approx 3.5°$ w.r.t. 0 V/Å) can be observed, and no stable droplet shape is observed beyond the 0.07 V/Å. Eventually, water droplets tend to change into a roughly conical shape at E = 0.08 V/Å. Hence, 0.08 V/Å is the critical electric field potential at which droplet elongates in the field direction on the graphite surface; a similar trend of contact angle with increasing electric field strength was also reported by Daub et al. [35] and Kargar et al.[36] Interestingly, the authors reported the crucial part of the Van der Waals interaction among the water molecules and graphene wherein the electric field value needed to stretched the droplet must be greater than that expected from the classical Young–Laplace equation [41].

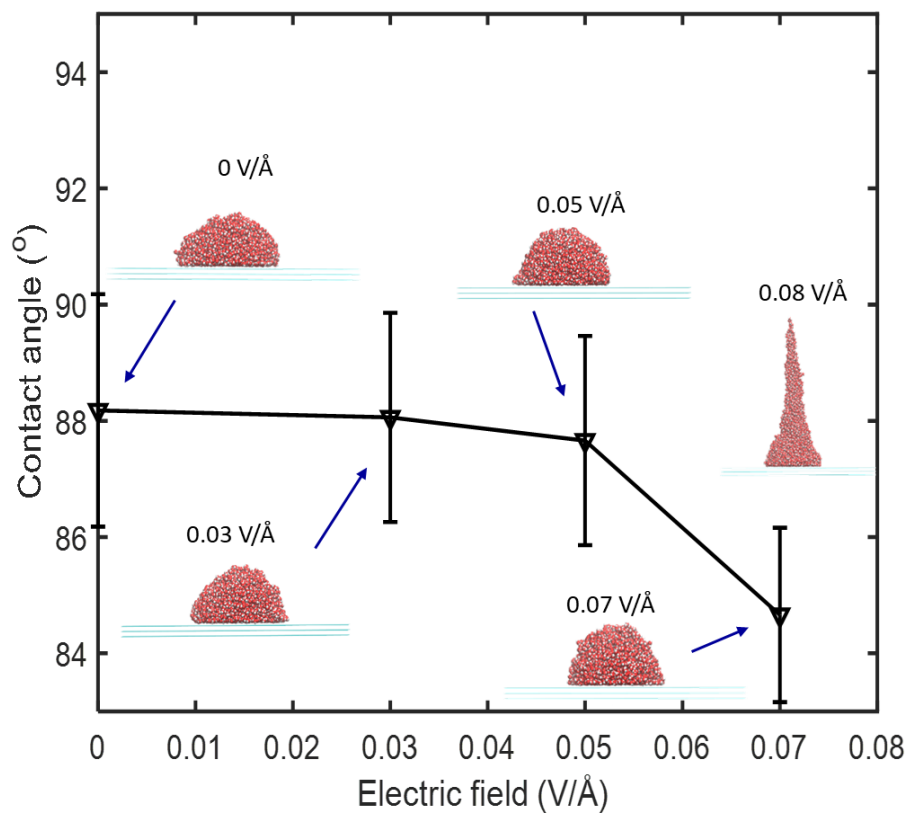

**Figure 1.** The contact angle of the water (3000 molecules) on the graphite sheet at different electric fields.

Further, the characteristics of the water nanodroplet under the effect of the electric field in the proximity of the graphite sheet is analyzed through the density profiles. In this regard, we have plotted the density profiles perpendicular to the surface (i.e., along the z-direction) for the selected cylindrical region (diameter 40 Å, yellow region) of the droplet as shown in **Figure 2. (please see the insets image of Figure2).** The axis of the cylinder is originated from the center of the mass of the droplet projected on the surface. This method is used to extract the actual mass density of ion pairs within the droplet along the z-coordinate. The atomic coordinate of the ion pairs present in the droplet is updated in the cylindrical regime while calculating the density profile. The bin width along the z-axis is taken as 0.5 Å, which is small enough to predict the density profile with reasonable accuracy. The density profile of the equilibrium droplet is calculated for 5 ns in the interval of 0.01 ps.

At 0 V/Å, near the graphite surface, two distinct density peaks can be detected at distances of 3.2 Å (first peak) and 6.2 Å (second peak) having peak heights of ~2.25 and ~1.28 $g/cm^3$, respectively which shows good agreement with data stated in the literature [37]. Interestingly, both peaks almost remain unchanged with the applied electric fields (see **Figure 2.** 0.05, and 0.07 V/Å) due to the substantial interaction within water and graphite sheets in the interfacial zone. However, the liquid-vapor interfacial region width increases with increasing the intensity of electrical field, as can be observed from (**Figure 2**); the density curve gets smooth and wider at the liquid-vapor interface ($\approx 25$Å). Furthermore, the inset image of the density contour reveals wider liquid-vapor width on the application of the electric field.

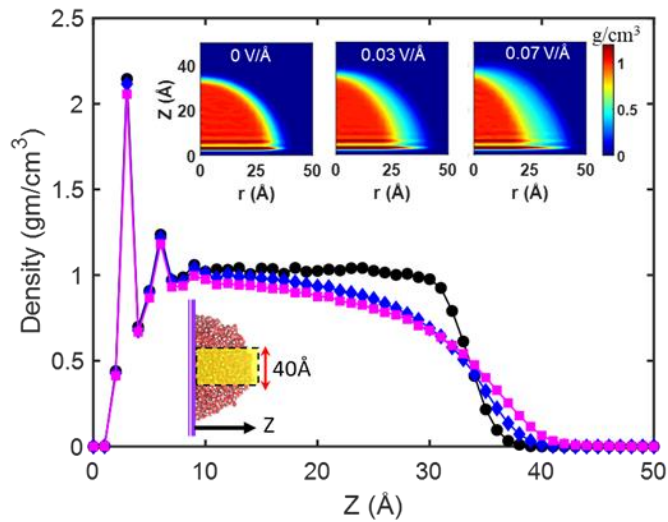

**Figure 2.** Density profiles of water through the centerline of the water droplet on graphite surface

## 3.2 Effect of Electric Field on Aqueous Hydrophilic IL

We now extend our study to explore the impact of the electric field on the aqueous droplet with various weight percentages of hydrophilic IL, such as [EMIM][BF$_4$]. Here, the equilibrated drop of aqueous [EMIM][BF$_4$] such as 10 wt.%, 30 wt.%, and 50 wt.% of IL on the graphite surface without electric (i.e., E0 V/Å) field are taken from our previous publication [32]. Variation of contact angle with the electric field is shown in (**Figure 3a,**) and the corresponding snapshot of droplet shapes are shown in **(Figure 3b).** Upon implementation of the lower electric field (0.01 to 0.05 V/Å), no significant deformation is observed in the droplet shape (see **Figure 3b**) and hence the contact angle remains almost the same. The equilibrium state of the droplet is nearly spherical cap-shaped (see **Figure 3b**). However, significant decrements in contact angle value, as well as shape deformation, are observed for above 0.07 V/Å. At the higher electric field of, i.e., 0.09 V/Å, the droplet elongated in the vertical direction for 10 wt. % and 30 wt. % of IL, and eventually elongated. Therefore, the critical electric field for 10 and 30 wt. % is 0.09 V/Å after which elongation in the shape of the droplet was observed. Nevertheless, for higher weight % of IL, such as 50 wt.%, and 70wt% the critical electric field observed is 0.1 V/Å. The reason behind these behaviors is the strong adhesive force towards the graphite due to the hydrophilic nature of [EMIM][BF$_4$].

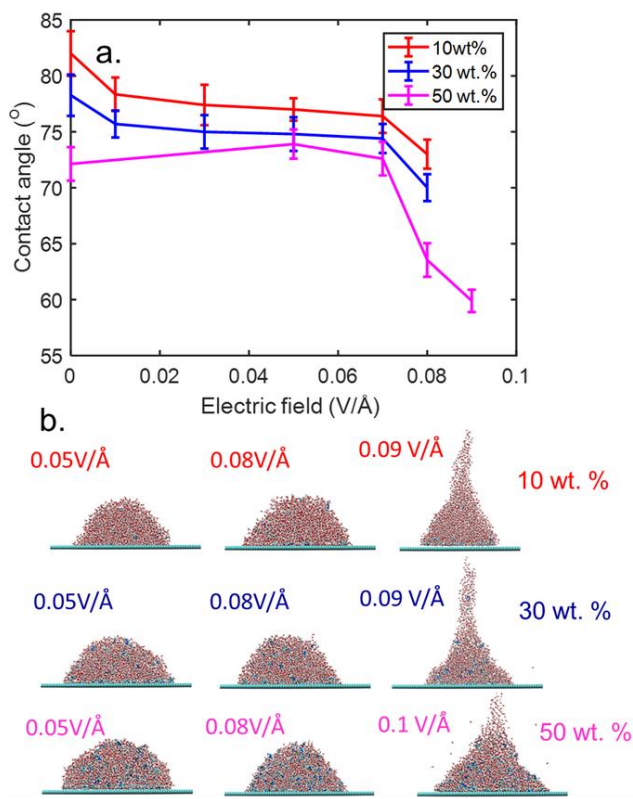

**Figure 3.** a) Contact angle of aqueous [EMIM] [BF$_4$] at the various electric field and b) Snapshot of aqueous [EMIM][BF$_4$] drop at the various electric field.

We further characterize the aqueous-IL droplet through 2D density contours. We analyzed the density distribution for IL and water in the drop separately, as shown in (**Figure 4** and **Figure 5),** respectively. We can observe from the contour of IL **(Figure 4),** for a low concentration of IL system such as 10 wt. %, IL molecules were not affected by the electric field; mostly, the water molecules move upward (as most of IL molecules are at the solid-fluid interface). Conversely, at higher concentrations (i.e., 30 wt. % and 50 wt. %), IL molecules are also pulled upward along with water molecules, as can be seen in **Figure 4** and **Figure 5.** However, the IL molecules in the vicinity of the graphite sheet don't get affected by the electric field due to the intense interaction energy between the sheet and IL.

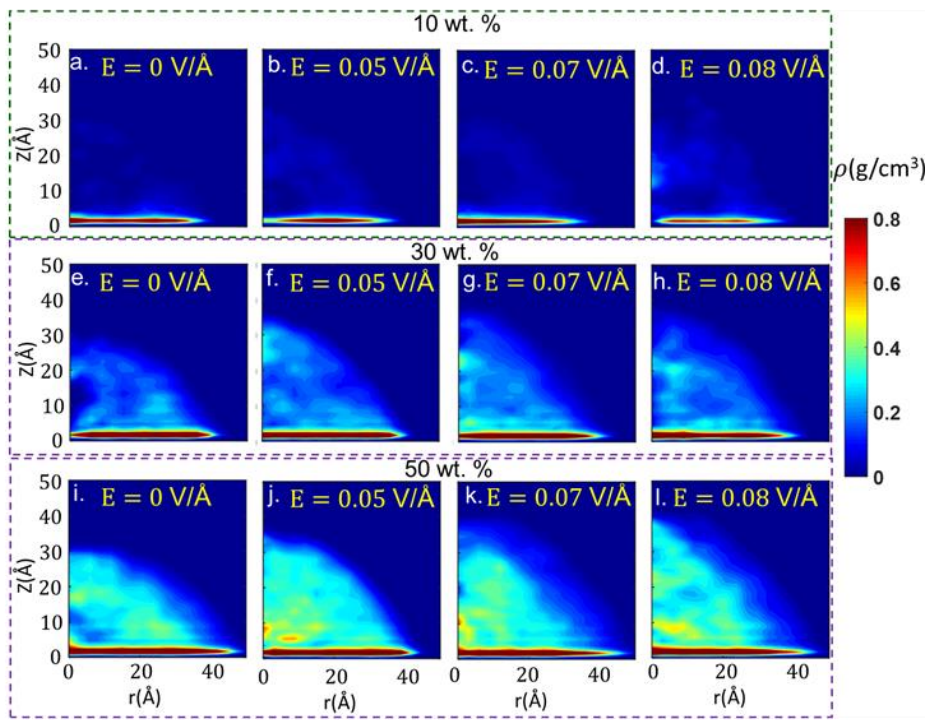

**Figure 4.** 2D density contours of the IL density distributions for aqueous [EMIM][BF$_4$] IL droplets

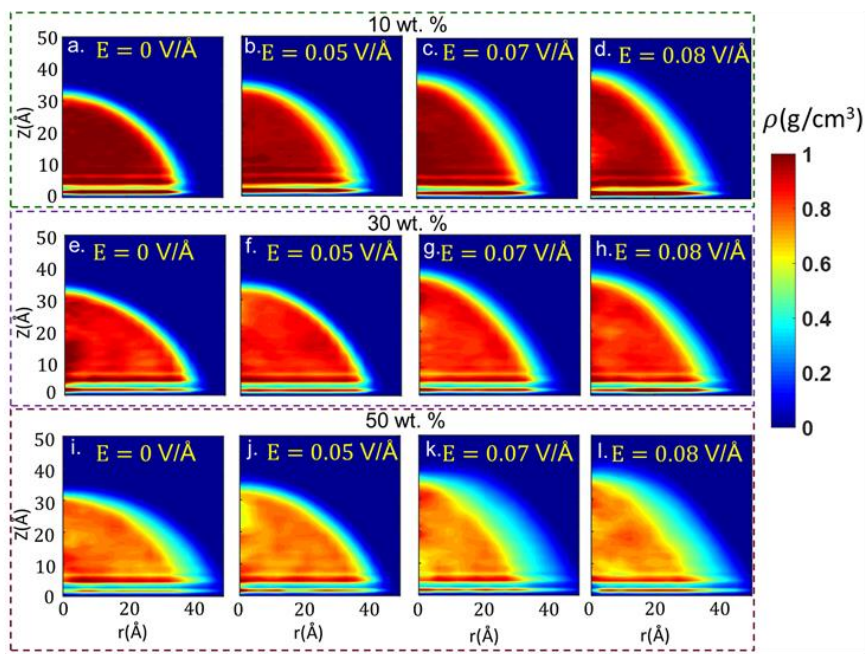

**Figure 5.** 2D density contours of water density distributions for the aqueous [EMIM][BF$_4$] IL droplet.

## 3.3 Effect of Electric Field on Aqueous Hydrophobic IL

We have so far discussed the effect of the electric field on aqueous hydrophilic IL. We have now examined the influence of the electric field on the aqueous hydrophobic IL, such as [EMIM][NTF$_2$]. Here, the contact angle of the droplet slightly decreases or unchanged with an increase in the strength of the electric field, similar to [EMIM][BF$_4$] However, in the case of hydrophobic [EMIM][NTF$_2$], the critical electric field above which droplet deformation observed is 0.08 V/Å for all concentration of IL molecules which is lower than that of aqueous hydrophilic ILs and pure water droplet as shown in **Figure 6.**

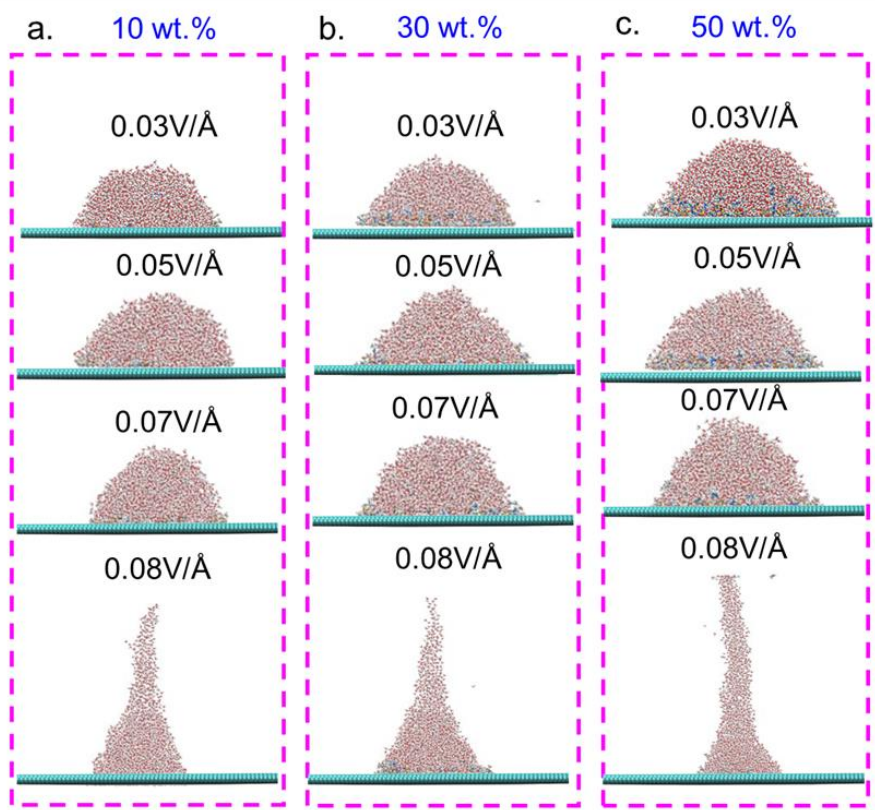

**Figure 6.** Snapshot of drop aqueous [EMIM][NTF$_2$] at the various electric field

Further, it can be observed from the density distribution contours of water [EMIM][NTF$_2$] and IL (see **Figure 7** and **Figure 8**) that IL molecules do not get affected by the electric field at all at lower concentration (10 wt.%), whereas water molecules are pulled upward. At the lower electric field, molecules of IL were found to be on a three-phase contact line. However, at higher electric filed (i.e., at 0.07 V/A), water molecules are completely pulled upwards; as a result, IL spread toward the center across the solid-fluid interface which is more visible for 30 wt. % of IL **(as shown in Figure 7.** snapshot of three phase contact line for 30 wt.%). Interestingly, at higher wt. % and electric field water molecules completely levitated on the IL, as can be seen in (**Figure 9i** and **Figure 10i**.)

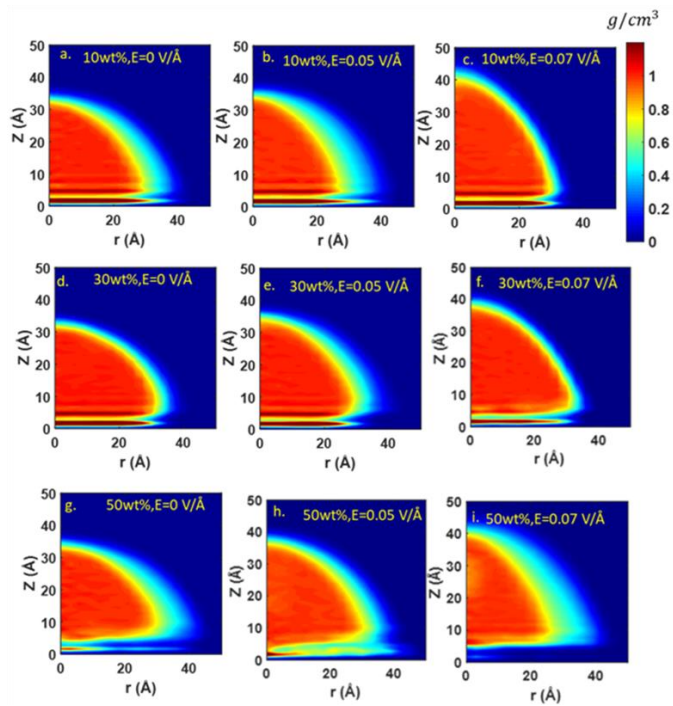

**Figure 7.** 2D density contours of the water density distributions for aqueous [EMIM][NTF$_2$]] IL droplets

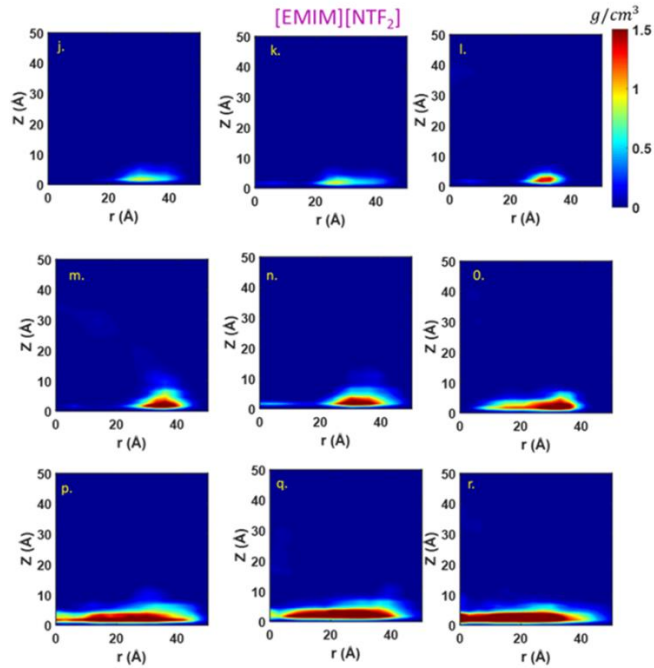

**Figure 8.** 2D density contours of the ILs density distributions for aqueous [EMIM][NTF$_2$] IL droplets

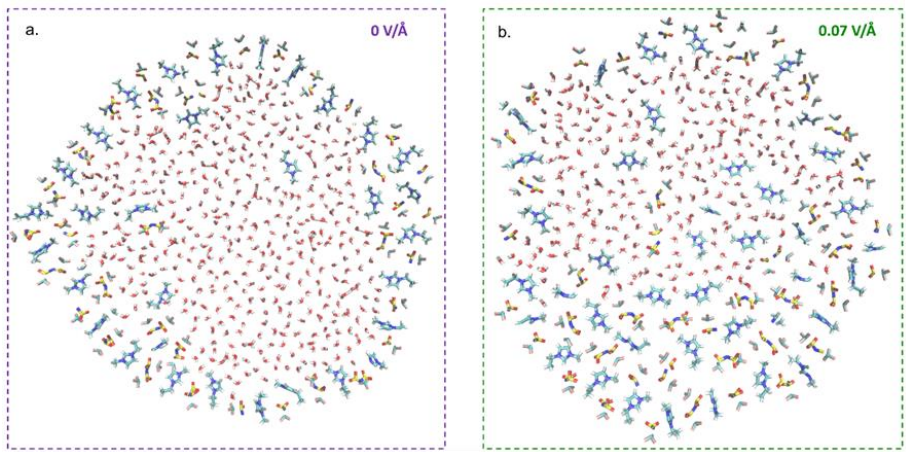

**Figure 9.** Snapshot of the first layer (near the surface) of aqueous 30 wt.% [EMIM][NTF$_2$] ILs droplet with a) E=0 V/Å and b) 0.07 V/Å, representing the distribution and orientation of molecules near graphene surface at three-phase contact line

### 3.4 Effect of Negative Electric Field on the Aqueous-IL Droplet

We have discussed the effect of a positive electric field, i.e., the electric field applied in an upward direction. However, the significant impact of the negative electric field on the contact angle was also reported in the literature[38]. The contact angle on the negative surface is found somewhat lower than that on the positively charged surface as observed by Dong et al.[20] for nanodroplet of [EMIM][NTF$_2$] on graphene sheets. Daub et al. [35]. reported that the positive electric field on 2000 SPC/E water molecules at 0.03 V/Å, induce lower contact angle (84.2±3.9°) than the negative electrical field (89.9 ±1.5°). This exhibits an inclination opposite molecular alignment since oxygen atom points away from the solid surface. Yen et al.[39] found that the negative field has a stronger electro stretching influence analogous to the previous study[35], which was attributed to the bias polarity of the water molecule under the electric field. Song et al. [25] indicate the considerable effect of the electric field direction on the wettability of [BMIM][BF$_4$] nano-drop on silicon substrate. It has been found that the ionic liquid droplet displays better wetting behavior in the negative electric field compared positive electric field, and accumulations of cations and anions in the first wetting layer found responsible for contributing to the spreading of an ionic liquid droplet on the solid substrate and cause the contact angle to decrease. Here, Here, (**Figure. 10),** we have shown 30 wt.% aqueous droplets of [EMIM][BF$_4$] and [EMIM][NTF$_2$] with negative electric field. However, we do not find any significant change for the negative electric field.

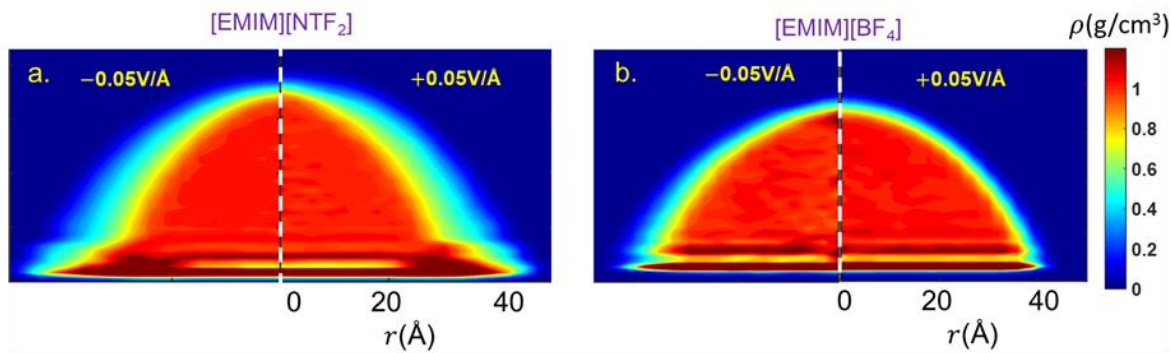

**Figure 10.** Density profiles of 30wt% a.) [EMIM][NTF$_2$]  b) [EMIM][BF$_4$] $at \pm 0.05$ V/Å

**3.5 Hydrogen Bond Analysis**

The hydrogen bond between water molecules is significantly susceptible to the electric field, which is exhibited by the increase of average hydrogen bond (HB) per molecule with the rise of the external electric field strength. Ren et al. [37] reported the impact of the electric field on the formation of HBs. The preferred orientation of water dipoles is along the electric field direction resulting hydrogen atoms to moves away from the surface increasing availability of water molecules for hydrogen bonding that results in the increase of HBs at the higher magnitude of the external electric field. Similarly, Daub et al. reported [35] the notable effect of the electric field on the alignment of water molecules, generally affecting the hydrogen bonding between surface molecules.

Song et al.[40] reported water molecules to reorient due to the rivaling force within electric field and intermolecular force. During the alteration, hydrogen bonds regenerate which could affect wetting behaviors. The direction of the dipole inclines to follow the direction of the electric field. They are disorderly distributed under the field strength is less than 0.8 V/nm, therefore the hydrogen bonds nearly remains unchanged. However, when the electric field exceeds a critical value the average number of hydrogen bonds decreases abruptly. This is due to the electric field strength is intensely sufficient that the disordered dipole distribution turned into ordered, and the hydrogen bond nets are destroyed. As a result, the average number of hydrogen bonds decreases speedily. The average number of HBs is largest when a droplet wets a flat silicon surface and smallest on a ramp-shaped surface.

Correspondingly, we have also estimated HB between anion-water and water-water to understand the nature of droplets on the application of electric field for 50 wt.% of [EMIM][BF$_4$] and [EMIM][NTF$_2$]. The HBs between anion- water and water-water has been determined based on geometric criteria by Koishi [41] shown in **Figure 11.**

**Figure 12.** shows the water-water and anion-water HBs for 50 wt. % [EMIM][BF$_4$] for the different electric field parameters. It can be observed that the average number of HBs between water - water decreases with an increase in the electric field. Conversely, the overall anion-water HBs are constant w.r.t electric filed (i.e.$\langle HB \rangle \approx 0.26$). The anion- water HB originates at the solid-liquid interface and then propagates to the droplet bulk with the increase in the electric field, as shown in **Figure 12.**

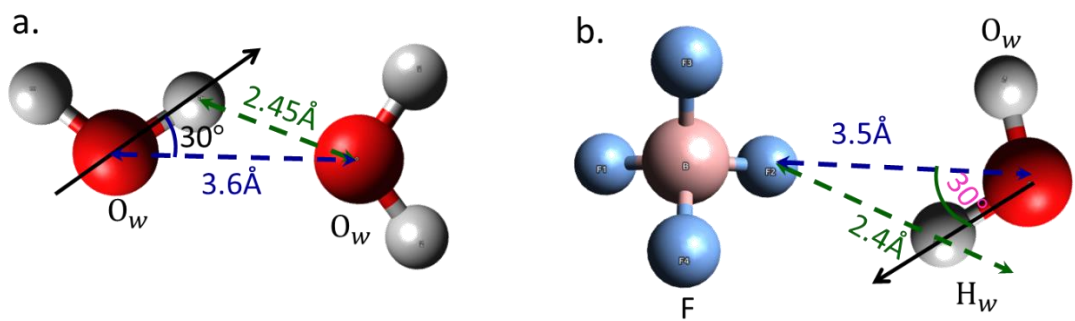

**Figure 11.** Hydrogen bond (HB) geometric criteria (a) for water-water (b) for anion-water.

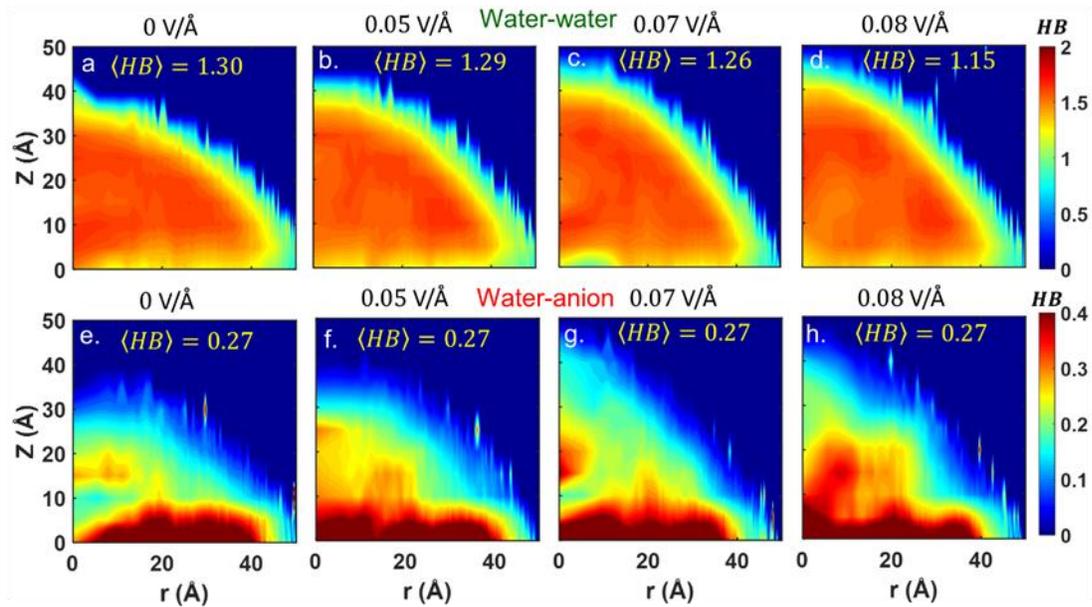

**Figure 12.** Average numbers of HBs between water-water and between water-anion of 50 wt. % aqueous [EMIM][BF$_4$] droplet for different electric field strength.

On the other hand, in the hydrophobic IL system, as shown in (**Figure 13**)**,** water-water HBs decreasing, whereas anion-water HBs increases with an increase in the electric field. Anion–water hydrogen bonding is mostly observed at the three-phase contact line and propagated into the center of the droplet across solid-fluid interface while increasing the electric field, i.e., at 0.07 V/A, as shown in **Figure 13.** Because of the hydrophobic nature of the [NTF$_2$] anions, [EMIM][NTF$_2$] IL molecules are not preferred in the bulk of the droplet. However, the preferential adsorption of [EMIM][NTF$_2$] at the three-phase contact line as well as at the solid-fluid interface (at higher potential) results from the ability of the [NTF$_2$] anions to form hydrogen bonds in those regions.

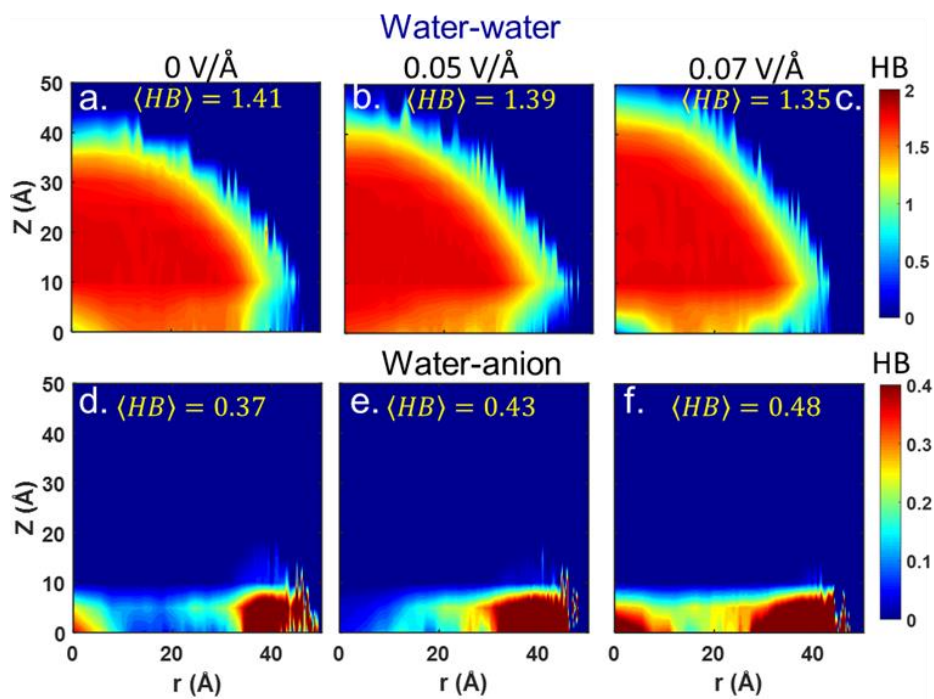

**Figure 13.** Average numbers of HBs between water-water and between water-anion of 50wt% aqueous [EMIM][NTF$_2$] droplet for different weight percentages.

Further, we have plotted Hydrogen-Bond profile normal to the surface to compare the HB properties in the surface region with those in the droplet bulk **(Figure. 14),** which shows near the surface hydrogen-bond between water-water is less compared to the bulk of the droplet whereas the HB between anion-water is more near the surface compared to the bulk of the droplet which is more pronounced at higher concentration. This observation suggests that the anion-water hydrogen bonding is more preferable compared to the water-water hydrogen bonding near the surface.

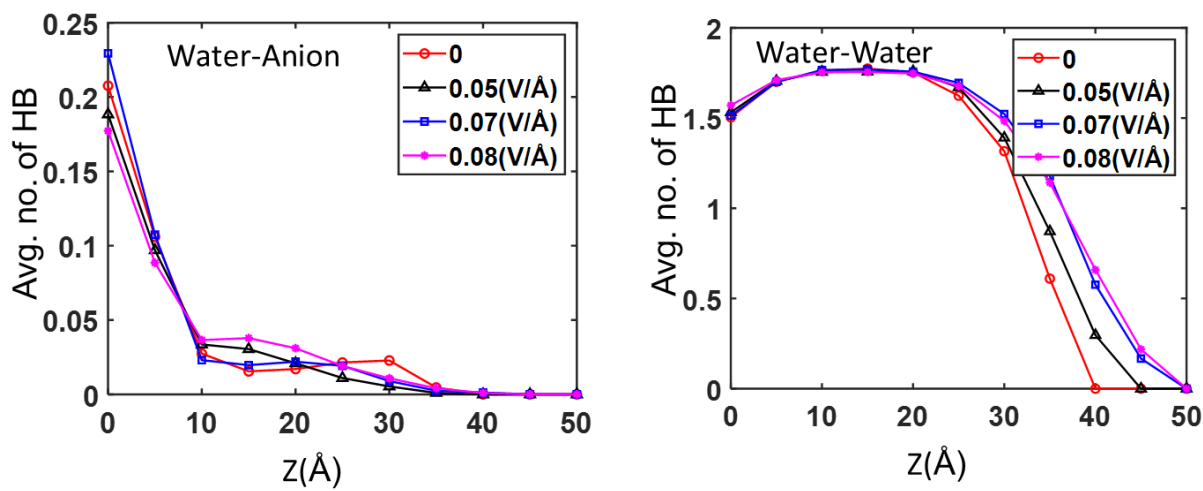

**Figure 14.** Hydrogen bond profile of droplets along the direction normal to the surface (i.e., the z-direction)

**Conclusion:**

We have first investigated the wetting behavior of pure water (SPC/E) droplets on the graphite sheet under the effect of the external electrical field. The contact angle of the droplet is found unchanged up to 0.05 V/Å. However, at 0.07 V/Å, a noticeable decrement of contact angle ($\approx 3.5°$ w.r.t. 0 V/Å) can be observed, and no stable droplet shape is observed beyond the 0.07 V/Å. Eventually, water droplets tend to change into a roughly conical shape at E = 0.08 V/Å. Hence, 0.08 V/Å is the critical electric field potential for pure water droplet. The z-density profile shows two pronounced density peaks, can be identified at distances of 3.2 Å (first peak) and 6.2 Å (second peak) with peak heights of ~2.25 and ~1.28 g/cm$^3$, respectively which shows good agreement with data reported in the literature [37]. Interestingly, both peaks almost remain unchanged with the applied electric fields due to the strong interaction between water and graphite sheet in the interfacial region. Subsequently, we extend our study to explore the effect of the electric field on the aqueous droplet with various weight percentages of hydrophilic IL, such as [EMIM][BF$_4$] and hydrophobic IL [EMIM][NFT$_2$]. In case of [EMIM][BF$_4$], upon application of the lower electric field (0.01 to 0.05 V/Å), no significant deformation is observed in the droplet shape and hence the contact angle remains almost same. The equilibrium state of the droplet is nearly in spherical cap-shaped. However, significant decrements in contact angle value, as well as shape deformation, are observed for above 0.07 V/Å. At the higher electric field of, i.e., 0.09 V/Å, the droplet is stretched longer in the vertical direction for 10 wt. % and 30 wt. % of IL, and eventually elongated. Therefore, the critical electric field for 10 and 30 wt. % is 0.09 V/Å after which elongation in the shape of droplet was observed. Nevertheless, for higher weight % of IL, such as 50 wt.%, and 70wt% the critical electric field observed is 0.1 V/Å. The average number of HBs between water - water decreases with an increase in the electric field. Conversely, the overall anion-water HBs are constant w.r.t electric filed (i.e.⟨HB⟩ ≈ 0.26). The anion- water HB originates at the solid-liquid interface and then propagates to the droplet bulk with the increase in the electric field. However, in the case of hydrophobic IL [EMIM][NTF$_2$], the critical electric field above which droplet deformation observed is 0.08 V/Å for all concentration of IL molecules which is lower than that of aqueous hydrophilic ILs. At the lower electric field, molecules of IL were found to be on a three-phase contact line. However, at higher electric filed (i.e., at 0.07 V/A), water molecules are completely pulled upwards; as a result, IL spread toward the center across the solid-fluid interface which is more visible for 30 wt. % of IL. Hydrogen bond analysis shows that water-water HBs decreases, whereas anion-water HBs increases with an increase in the electric field. Anion–water hydrogen bonding is mostly observed at the three-phase contact line and propagated into the center of the droplet across solid-fluid interface while increasing the electric field. We also examine the effect of negative electric field on the wetting behavior of aqueous IL droplet and do not find any significant change on the droplet behavior compared to that of positive electric field.

**Acknowledgments**

We would like to acknowledge the Department of Chemical and Biochemical at IIT Patna for providing computational resources for this research. This work is supported by grants from Science and Engineering Research Board (SERB), Government of India (Project File no. ECR/2018/002600). We also would like to acknowledge the Center for Development of Advanced Computing (C-DAC), India for generous allocation of computing resources.